\documentclass[prb,twocolumn,amsmath,amssymb,superscriptaddress,floatfix,nofootinbib,aps,10pt]{revtex4-2}

\usepackage{graphicx}
\usepackage{dcolumn}
\usepackage{bm}
\usepackage{ textcomp }
\usepackage{color}
\usepackage{amsmath}
\usepackage{amssymb}
\usepackage{xcolor}
\usepackage[colorlinks = true,
            linkcolor = blue,
            urlcolor  = red,
            citecolor = blue,
            anchorcolor = blue]{hyperref}
\usepackage{easyReview}
\usepackage{mathdots}
\usepackage{float}
\usepackage{lineno}
\usepackage{todonotes}
\usepackage{url}
\usepackage{array}
\usepackage{todonotes}
\usepackage{xcolor,colortbl}

\usepackage{lipsum, babel}

\definecolor{LightBlue}{rgb}{0.8,0.8,0.8}

\usepackage[normalem]{ulem}

\begin{document}
\title{Coupled Hubbard ladders at weak coupling: Pairing and spin excitations}
\date{\today}

\author{Thomas A. Maier}
\affiliation{Computational Sciences and Engineering Division, Oak Ridge National Laboratory, Oak Ridge, Tennessee 37831-6164, USA}

\author{Elbio Dagotto}
\affiliation{Department of Physics and Astronomy, The University of Tennessee, Knoxville, Tennessee 37966, USA}
\affiliation{Materials Science and Technology Division, Oak Ridge National Laboratory, Oak Ridge, Tennessee 37831, USA}

\begin{abstract}

The Hubbard model provides a simple framework in which one can study how certain aspects of the electronic structure of strongly interacting systems can be tuned to optimize the superconducting pairing correlations and how these changes affect the mechanisms giving rise to them. Here we use a weak-coupling random phase approximation to study a two-dimensional Hubbard model with a unidirectional modulation of the hopping amplitudes as the system evolves from the uniform square lattice to an array of weakly coupled two-leg ladders. We find that the pairing correlations retain their dominant $d_{x^2-y^2}$-wave like structure and that they are significantly enhanced for a slightly modulated lattice. This enhancement is traced backed to an increase in the strength of the spin-fluctuation pairing interacting due to favorable Fermi surface nesting in the modulated system. We then use a random-phase approximation BCS framework to examine the evolution of the neutron resonance in the superconducting state. We find that it changes only weakly for moderate modulations, but breaks up into two distinct resonances at incommensurate wave-vectors in the limit of weakly coupled ladders. 

\end{abstract}

\maketitle

\section{Introduction}
 
The Hubbard model \cite{Hubbard1963} provides a simple framework to study how a purely repulsive interaction between fermions can give rise to a superconducting instability. It has been extensively studied in the context of the cuprates, for which it was argued to provide an appropriate low-energy effective description of the electronic degrees of freedom \cite{Anderson1987,Zhang1988}. Advanced numerical approaches have been extensively used to study this model on square lattices~\cite{Dagotto:rmp94,Scalapino2012,Qin2021,Arovas2021}. In this case, the ground state of the doped model is still under debate~\cite{Maier2005a,Staar2014,Zheng2017,Qin2020,Jiang2020,Chung2020} and likely depends on details in the model parameters, such as the plaquette diagonal hopping $t'$ strength and sign. Nevertheless, several studies have found evidence that it exhibits properties remarkably similar to what is observed in the cuprates, including antiferromagnetism, pseudogap and superconducting behavior, as well as striped spin and charge density waves \cite{Scalapino2012,Qin2021,Arovas2021}. Related to these developments in square lattices, in the 90's it was predicted that two-leg ladders would have a spin gap, exponentially decaying spin correlations, and they would superconduct upon doping~\cite{riera,rice,Vuletic2006}. These predictions were experimentally confirmed~\cite{uehara96,pressure-piskunov,ladder3}. For this geometry, essentially exact calculations can be carried out, and the doped model is known to support a Luther-Emery liquid phase with power-law superconducting correlations \cite{Noack1995,Dolfi2015,Jiang2020b}. 

Unbiased and controlled numerical calculations are necessary to accurately and reliably predict the properties of the Hubbard model in the non-perturbative regime relevant for the cuprates, where the Coulomb repulsion between the electrons is of the same magnitude as their kinetic energy. On the other hand, weak-coupling random-phase approximation (RPA) based approaches, while perturbative, provide a level of simplicity and transparency, and therefore insight, that numerical methods cannot offer. Recent studies have shown that RPA-based weak coupling approaches describe certain behavior in Hubbard models remarkably well. For example, systematic studies of the leading pairing instabilities in a 2D square lattice Hubbard model with increasing Coulomb repulsion $U$ found that weak coupling RPA predictions agreed exceptionally well~\cite{Romer2020} with those of non-perturbative quantum Monte Carlo dynamic cluster approximation (DCA) \cite{Maier2005} results. Similarly, for a Hubbard model in a two-leg ladder geometry, RPA calculations within a fluctuation exchange approximation (FLEX) were shown to capture the main features of density matrix renormalization group (DMRG) results for the magnetic and charge dynamical response \cite{Nocera2018}. 

Here, we use RPA calculations to study the Hubbard model for geometries in between these two limits, i.e. for an array of coupled two-leg ladders. In particular, we are interested on how the superconducting pairing correlations change when the system evolves from the uniform square lattice limit to the two-leg ladder geometry, as the hopping amplitude between the ladders is reduced (see Fig.~\ref{fig:HM}). A very similar system with a unidirectional modulation of the hopping amplitudes was recently investigated with the density matrix renormalization group in the context of the question of whether striped charge density wave order can enhance the superconducting correlations \cite{jiang2021stripe}. This study found an order of magnitude enhancement of the superconducting correlations even for modest modulations for reasons that remain investigated in more detail. For the opposite limit of an isolated two-leg ladder, a similar enhancement was found in recent RPA-based FLEX calculations when the rung-to-leg ratio of the nearest neighbor hopping increases \cite{Sakamoto2020}. Understanding how superconductivity can be optimized by tuning certain aspects of the electronic structure is not only useful for the search for new materials with improved properties, but can also provide insight into the nature of the mechanisms responsible for the pairing correlations. Here, we find an enhancement of the superconducting pairing strength for a slightly modulated hopping amplitude, and show that within RPA, this enhancement arises from an increase in the strength of the spin-fluctuation pairing interaction. 

We then use an RPA/BCS framework to study how the superconducting gap affects the spin-fluctuation spectrum under variation of the inter-ladder coupling. For the uniform square lattice, this formalism is known to provide a framework in which one can understand the neutron resonance observed in the superconducting state of the cuprates \cite{Norman2007,Scalapino2012}. Here, we examine how this neutron resonance and its dispersion evolve as the inter-ladder coupling decreases towards the isolated two-leg ladder limit. We find that the resonance becomes stronger for the slightly modulated case with the largest pairing correlations. For weakly coupled ladders, the resonance is still present, albeit at a different wavevector.



\section{Model and Method}

\begin{figure}[ht]
    \centering
    \includegraphics[width=0.35\textwidth]{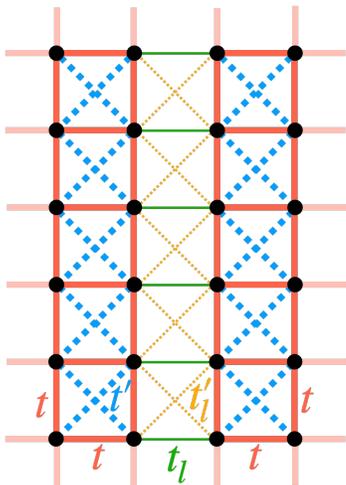}
    \caption{Illustration of the Hubbard model on a square lattice with modulated hopping amplitudes. Here we consider nearest-neighbor hopping within the two-leg ladders ($t$) that run along the $y$-direction and between the ladders ($t_l=rt$), as well as next nearest neighbor hopping within ($t'$) and between the ladders ($t'_l=rt'$). The parameter $r$ specifies the ratio between intra- and inter-ladder hopping amplitudes. \label{fig:HM}} 
\end{figure}

We consider a two-dimensional Hubbard model on a square lattice
\begin{equation}\label{eq:HM}
    {\cal H} = \sum_{ij, \sigma} t_{ij} c^\dagger_{i\sigma}c^{\phantom\dagger}_{j\sigma} + U\sum_i n_{i\uparrow}n_{i\downarrow}\,,
\end{equation}
 with modulated hopping amplitudes along the $x$-direction. Here, $c^{\dagger}_{i\sigma}$ ($c^{\phantom\dagger}_{i\sigma}$) creates (annihilates) an electron on site $i$ with spin $\sigma$, $n_{i\sigma}=c^\dagger_{i\sigma}c^{\phantom\dagger}_{i\sigma}$ is the electron number operator and $U$ is the on-site Coulomb repulsion. As illustrated in Fig.~\ref{fig:HM}, the system can be considered as an array of coupled two-leg Hubbard ladders that run along the $y$-direction. We set the hopping amplitudes within the ladders to $t_{ij} = -t$ for $i,j$ nearest neighbors (along the rungs) and to $t_{ij}=-t'$ when $i$ and $j$ are next-nearest neighbors along the plaquette diagonals. The hopping between the ladders is set to $t_l = rt$ for nearest-neighbor sites and $t'_l=rt'$ for next-nearest neighbor sites. Here, $r$ is a tunable parameter that controls the relative amplitudes between intra- and inter-ladder hoppings. The Hamiltonian in Eq.~\eqref{eq:HM} reduces to the usual uniform 2D Hubbard model for $r=1$. For $r\lesssim 1$, it describes a Hubbard model with modulated hopping amplitudes along the $x$-direction and uniform amplitudes along the $y$-direction. In the opposite limit of very small $r\ll 1$, it describes an array of weakly coupled Hubbard ladders. In the following, we use $t=1$ as energy unit, and set $t'=-0.25t$. Throughout the manuscript, the electron density is fixed to $\langle n\rangle = 0.85$.

We use a random-phase approximation framework to study this model. In the non-uniform case, $r\neq 1$, there are two sites in the unit cell (left and right legs of the ladders), and the model can be considered a two-orbital model, where the two orbitals correspond to the two sites. In the multi-orbital RPA framework \cite{Kubo2007,Graser2009}, the pairing vertex $\Gamma_{\ell_1\ell_2\ell_3\ell_4}(k,k')$ for scattering a singlet pair $(k\uparrow\ell_1, -k\downarrow\ell_4)$ in orbitals (leg 1 or 2) $\ell_1$ and $\ell_4$ to a pair $(k'\uparrow\ell_2, -k'\downarrow\ell_3)$ in orbitals $\ell_2$ and $\ell_3$ is given by
\begin{eqnarray} \label{eq:Gamma}
\Gamma_{\ell_1\ell_2\ell_3\ell_4}(k,k')&=&\biggl[\frac{3}{2}\ {\cal U}\chi_s^{\rm
RPA}(k-k', 0){\cal U}\\ \nonumber &-&\frac{1}{2}\ {\cal U}\chi_c^{\rm RPA}(k-k', 0){\cal U}\\
\nonumber &+&{\cal U})\biggr]_{\ell_1\ell_2\ell_3\ell_4}.
\end{eqnarray} 
Here ${\cal U}$ is a $4\times 4$ matrix with elements ${\cal U}_{\ell_1\ell_2\ell_3\ell_4}=U\delta_{\ell_1\ell_2}\delta_{\ell_3\ell_4}\delta_{\ell_1\ell_4}$. 
The RPA spin $\chi^{\rm RPA}_s$ and charge $\chi^{\rm RPA}_c$ susceptibilities
are
\begin{eqnarray}\label{eq:chi}
  \chi^{\rm RPA}_s(q,i\omega_m)&=&\chi^0(q,i\omega_m)[1-{\cal U}\chi^0(q,i\omega_m)]^{-1} \\ \nonumber
  \chi^{\rm RPA}_c(q,i\omega_m)&=&\chi^0(q,i\omega_m)[1+{\cal U}\chi^0(q,i\omega_m)]^{-1},
\end{eqnarray}
where $\omega_m=2m\pi T$ is a bosonic Matsubara frequency and the matrix elements of the bare susceptibility $\chi^0(q)$ are given by
\begin{eqnarray}\label{eq:chi0}
  \chi^0_{\ell_1\ell_2\ell_3\ell_4}(q,i\omega_m)&&=-\frac{T}{N}\sum_{k,\omega_n}
  G^0_{\ell_3\ell_1}(k+q,i\omega_n+i\omega_m) \\\nonumber
  &&\,\,\,\;\;\times \,G^0_{\ell_2\ell_4}(k,i\omega_n),
\end{eqnarray}
with $G^0_{\ell\ell'}(k,i\omega_n)$ the bare Green's function
\begin{equation}
  G^0_{\ell\ell'}(k,i\omega_n)=\sum_\mu\frac{a^\ell_\mu(k)a^{\ell'*}_\mu(k)}
  {i\omega_n-\xi_\mu(k)}\,
  \label{eq:4}
\end{equation}
and $\omega_n=(2n+1)\pi T$ a fermionic Matsubara frequency. 
The band energies ($\mu=\pm$) are 
\begin{eqnarray}
    \xi_{\pm}(k) = -2t\cos k_y \mp \zeta(k),
\end{eqnarray}
for the bonding (+) and anti-bonding (-) bands with 
\begin{eqnarray}
    \zeta(k) &=& \biggr[ t^2+t_l^2+2tt_l\cos 2k_x  \\\nonumber
               &+& 4\cos k_y[tt'+t_lt'_l+(tt'_l+t't_l)]\cos 2k_x\\\nonumber
               &+&  4\cos^2k_y({t^\prime}^2+{t^\prime_l}^2 +2t't'_l\cos 2k_x) \biggr]^{1/2}\,.
\end{eqnarray}
The orbital matrix elements $a_\pm^\ell$ are given by
\begin{eqnarray}
    a_+^1(k) &=& a_-^1(k)  = \frac{\zeta(k)}{\sqrt{\zeta^2(k)+|\epsilon_{12}(k)|^2}}\\\nonumber
    a_+^2(k) &=& -a_-^2(k) = \frac{\epsilon^*_{12}(k)}{\sqrt{\zeta^2(k)+|\epsilon_{12}(k)|^2}}\,,
\end{eqnarray}
with
\begin{equation}
    \epsilon_{12}(k) = -te^{ik_x}-t_le^{-ik_x}-2\cos k_y(t'e^{ik_x}+t'_le^{-ik_x})\,.
\end{equation}
Carrying out the usual analytic continuation to the real frequency axis, the dynamic spin susceptibility is given by
\begin{equation}\label{eq:dynamicSpinSus}
    \chi^{\rm RPA}_s(q,\omega) = \sum_{\ell_1,\ell_2} \chi^{\rm RPA}_{s, \ell_1\ell_1\ell_2\ell_2}(q, i\omega_m\rightarrow \omega+i\delta)\,.
\end{equation}

In terms of the scattering vertices, the pairing strength is given by the
leading eigenvalue $\lambda_\alpha$ of
\begin{equation} \label{eq:BSE}
    -\sum_j\oint\frac{dk'_{\parallel}}{2\pi
  v_{F_j}(k'_{\parallel})}
    \Gamma_{ij}(k,k')g^\alpha_j(k')=\lambda_\alpha g^\alpha_i(k)
\end{equation} with
\begin{eqnarray} \label{eq:Gamma2}
  \Gamma_{ij}(k,k')&=&\sum_{\ell_1\ell_2\ell_3\ell_4}a^{\ell_1}_{\nu_i}(k)a^{\ell_4}_{\nu_i}(-k)
    \Gamma_{\ell_1\ell_2\ell_3\ell_4}(k,k')\\ \nonumber
    &\times&a^{\ell_2*}_{\nu_j}(k')a^{\ell_3*}_{\nu_j}(-k').
\end{eqnarray} Here $j$ sums over the Fermi surfaces,
$v_{F_j}(k'_{\parallel})$ is the Fermi velocity $|\nabla_k\xi_{\nu_j}(k)|$ and
the integral runs over the Fermi surface. We generally find that for all values of $0<r<1$, the leading eigenvalue $\lambda_d(k)$ has mostly $d_{x^2-y^2}$ character. Note that for $r<1$, the C4 rotational symmetry of the square lattice is broken, and mixing with an $s$-wave component is expected. 

Before moving into our main results, what value of $r$ is realistic in our definition of hoppings via
expressions such as $t_l = rt$?
From inelastic neutron scattering (INS) experiments, estimations of how robust the ladder couplings
are can be made. For example, INS studies of large single crystals of La$_4$Sr$_{10}$Cu$_{24}$O$_{41}$~\cite{notbohm}, containing
copper two-leg ladders in its atomic structure, provided approximate values for the Heisenberg superexchange 
along legs $J_{leg}=186$~meV, rungs $J_{rung}=124$~meV, and in between ladders $J'=36$~meV. For the latter the calculation was done within RPA, see
footnote 19 in Ref.~\cite{notbohm}. This value is considered to be more uncertain than the rest according
to Ref.~\cite{notbohm}). The theory was done using a continuous unitary transformation. Another INS
study~\cite{matsuda}, this time for La$_6$Ca$_8$Cu$_{24}$O$_{41}$, reported that for their single crystals 
the values were $J_{leg}=J_{rung}=110$~meV, using for the theory component exact diagonalization of small clusters.
No results for the interladder coupling were given.
Similar numbers were obtained via optical conductivity experiments~\cite{nunner} as well. The common factor of all these
studies is that (a) the rung and ladder superexchanges are similar even among different materials and when using different techniques, (b) the cyclic four-spin interaction within a plaquette that appears in a higher order expansion around the atomic limit is small but not negligible (in our case we employ directly the Hubbard model and this plaquette-related coupling is already taken into account), and (c) the largest uncertainty is in the coupling between ladders, widely considered to be small because it involves a 90$^{o}$ degree Cu-O-Cu bond. Due to the uncertainty in point (c), for a crude estimation of the interladder coupling we will use the only value available $J'=36$~meV~\cite{notbohm}. With regards to the leg and rung
couplings an average of the values quoted above will be used, 
namely we employ $J_{leg}=J_{rung}=132$~meV. Assuming the canonical
relation between superexchange, hopping amplitude $t$ and Hubbard $U$, i.e. $J=4t^2/U$, we deduce
 a ratio $r=t_l/t = \sqrt{36/132}= 0.5$, well within the range $0.25<r<1$ to be discussed here. 
Because observables presented below appear to converge starting at $r=0.5$ and for smaller $r$, we expect results at the realistic $r$, likely smaller than the 0.5 previously discussed, to be approximately the same.

%
%
%
\section{Results}

\begin{figure}[ht]
    \centering
    \includegraphics[width=0.5\textwidth]{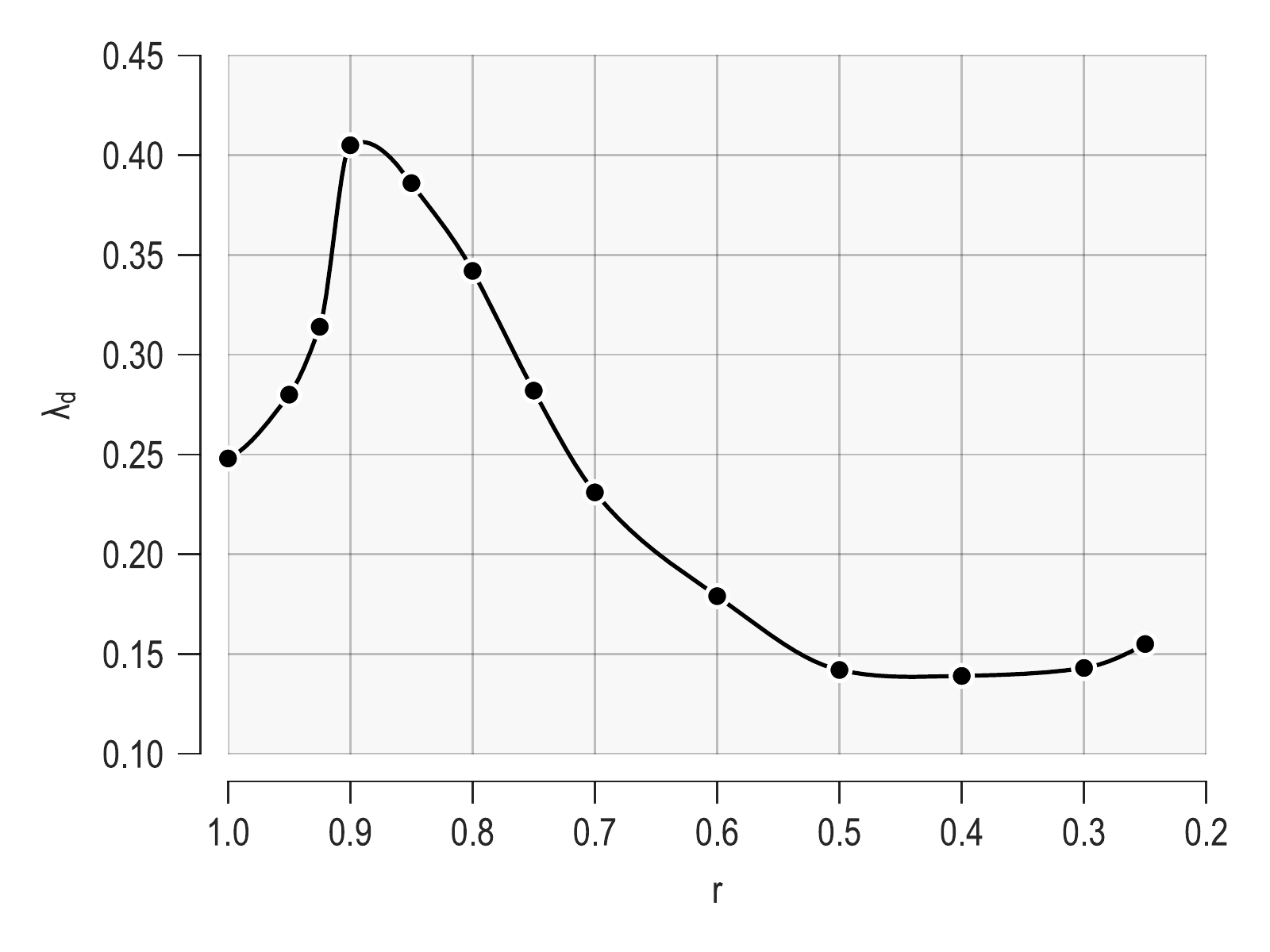}
    \caption{ The $d_{x^2-y^2}$-wave pairing strength $\lambda_d$ versus hopping ratio $r$ for a Hubbard model with filling $\langle n\rangle=0.85$ and Coulomb interaction $U=1.5t$. $\lambda_d$ is optimized for $r<1$ corresponding to a model with modulated hopping integrals along the $x$-direction, i.e. effectively an array of coupled two-leg ladders.  \label{fig:lambda_vs_r}} 
\end{figure}

We start by examining how the the leading $d$-wave eigenvalue $\lambda_d$ of Eq.~\eqref{eq:BSE} depends on the ratio $r$ between intra- and inter-ladder hopping. The Coulomb repulsion $U=1.5t$ is chosen so that $\lambda_d\lesssim 0.5$ in the RPA treatment. Figure~\ref{fig:lambda_vs_r} shows the RPA results for $\lambda_d(r)$ for $0.25<r<1$. Remarkably, we find that the maximum in $\lambda_d(r)$ does not occur for the uniform $r=1$ limit. Rather, $\lambda_d(r)$ displays a rapid initial rise as $r$ is reduced from $r=1$, and has a maximum for $r\approx 0.9$, before dropping again to smaller values for $0.9>r\gtrsim 0.4$. The point $r\approx 0.9$, where the pairing strength is maximized, corresponds to a system in which the nearest and next-nearest neighbor hopping amplitudes are periodically modulated along the $x$-direction. As noted, a similar enhancement of the superconducting correlations for modest hopping amplitude modulations was also found in a recent density matrix renormalization group study \cite{jiang2021stripe} of a Hubbard model with similar hopping amplitude modulations. While the precise reason for this enhancement remains unclear, it was argued that the spin fluctuation spectrum in the modulated array of effective two-leg ladders is optimal for pairing (the undoped two-leg ladders have a spin gap and thus lack the low-energy fluctuations that are detrimental to superconductivity). 

\begin{figure*}[t]
    \centering
    \includegraphics[width=0.95\textwidth]{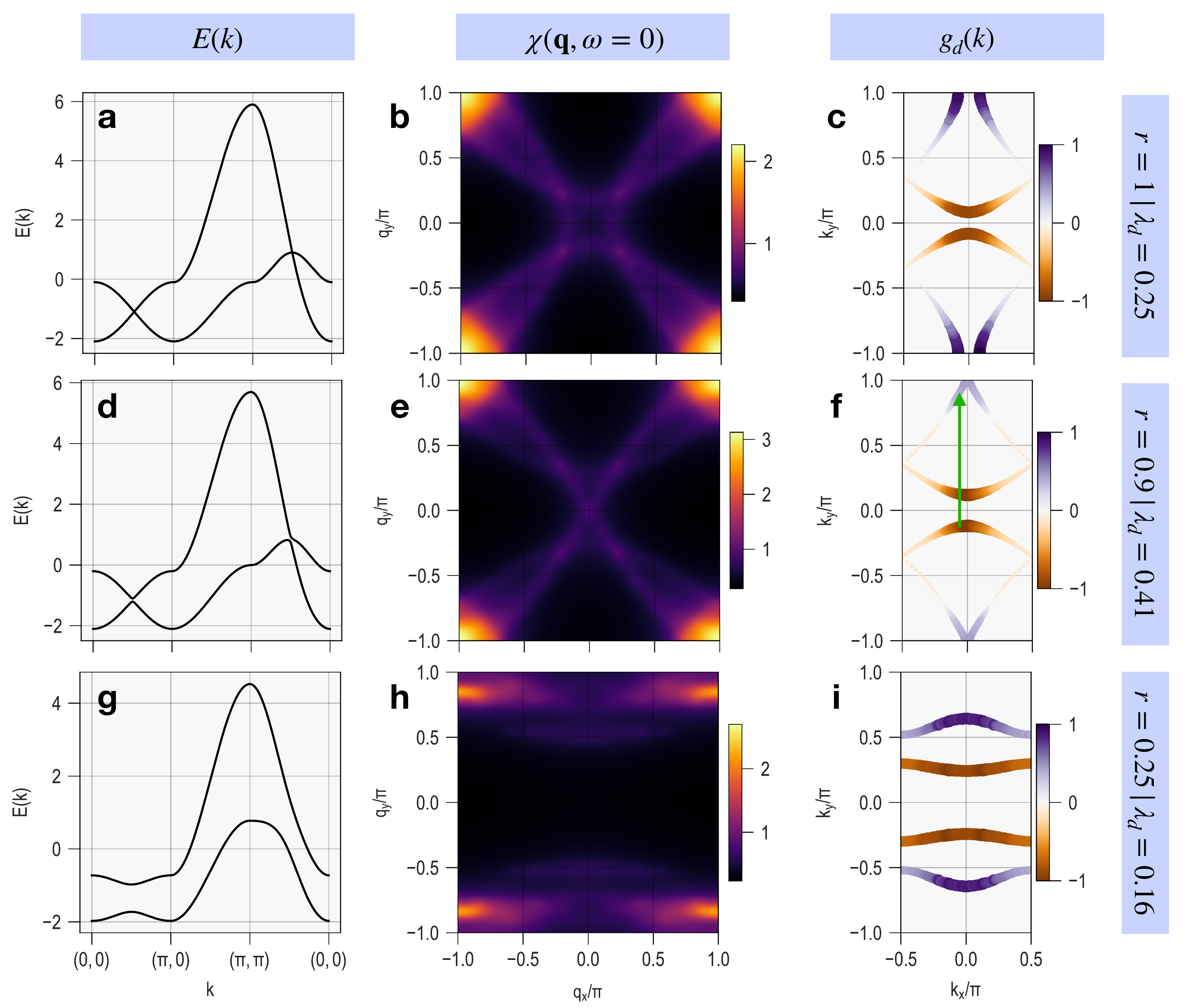}
    \caption{Bandstructure $E({\bf k})$ (left panels {\bf a}, {\bf d}, {\bf g}), static spin susceptibility $\chi({\bf q}, \omega=0)$ (middle panels {\bf b}, {\bf e}, {\bf h}) and leading eigenvector $g_d({\bf k})$ (right panels {\bf c}, {\bf f}, {\bf i}) for the Hubbard model with $U=1.5t$ and $\langle n\rangle=0.85$ for three different values of the hopping ratio $r$. As $r$ is reduced from 1 and the hopping is modulated, the response near ${\bf q}=(\pi,\pi)$ in $\chi({\bf q},0)$ increases first and then decreases, leading to the non-mononotic behavior observed in $\lambda_d(r)$ in Fig.~\ref{fig:lambda_vs_r}. The leading gap structure $g_d({\bf k})$ retains mostly $d_{x^2-y^2}$ character, but also aquires an $s$-wave contribution due to the breaking of the C4 lattice rotation symmetry. In the weakly coupled limit ($r=0.25$, panel {\bf i}), the gap is nodeless and switches sign between the bonding and antibonding Fermi surfaces. The green vector in panel {\bf f} corresponds to $q=(\pi, \pi)$ in the full Brillouin zone.\label{fig:Ek_chi_g}} 
\end{figure*}

To determine the origin of the enhanced pairing strength in our results, we now take a closer look at the variation of the electronic structure and spin susceptibility as the hopping ratio $r$ is reduced from one. Figure~\ref{fig:Ek_chi_g} shows the bandstructure $E(k)$ (left panels), static spin susceptibility $\chi_s^{\rm RPA}(q,\omega=0)$ (Eq.\eqref{eq:dynamicSpinSus}, middle panels) and $d$-wave eigenfunction $g_d(k)$ of Eq.~\eqref{eq:BSE} (right panels) for three different values of $r$ and the same parameters as used in Fig.~\ref{fig:lambda_vs_r}. 

The top row (panels {\bf a} - {\bf c}) shows the $r=1$ results for the uniform system. Even in this case, the bandstructure has two bands, due to the use of a two-site unit cell in the RPA treatment. The single band of the uniform model is simply folded along the boundaries $(\pi/2, k_y)$ and $(-\pi/2, k_y)$ of the reduced Brillouin zone of the two-site system, to give the two bands shown in Fig.~\ref{fig:Ek_chi_g}{\bf a}. The susceptibility $\chi^{\rm RPA}_s(q,0)$ in panel {\bf b} shows the largest intensity at $q=(\pi,\pi)$ as expected. The eigenfunction $g_d(k)$ in panel {\bf c} has the usual $d_{x^2-y^2}$ structure. Here the negative (brown) regions near $k=(0,0)$ originate from the anti-bonding band, which implicitly contains a shift of $k_x=\pi$, and therefore correspond to $k=(\pi,0)$ in the full Brillouin zone. These regions are simply folded back into the reduced Brillouin zone of the two-site system, for which $k=(0,0)$ is equivalent to $k=(\pi,0)$.

As $r$ is reduced to 0.9 (middle row panels {\bf d}-{\bf f}), a small gap opens in the bandstructure at the boundary of the reduced Brillouin zone at $k=(\pi/2,k_y)$. The spin susceptibility looses its C4 symmetry, but the peak with the highest intensity remains at $q=(\pi,\pi)$. In fact, the intensity of this peak is much higher, about 50\%, than for $r=1$. This explains the increase in $\lambda_d(r)$ for $r=0.9$, since, according to Eqs.~\eqref{eq:BSE} and \eqref{eq:Gamma}, the pairing strength is directly related to the magnitude of the zero frequency spin susceptibility $\chi_s^{\rm RPA}(q,\omega=0)$. This increase in $\chi_s^{\rm RPA}(q,\omega=0)$ can be traced to the change in the Fermi surface and improved nesting. The vertical green vector shown in panel {\bf f} is $q=(0,\pi)$ in the reduced Brillouin zone and corresponds to $q=(\pi,\pi)$ in the full Brillouin zone on which $\chi_s^{\rm RPA}(q,\omega=0)$ is evaluated. It illustrates that the regions on the Fermi surface that are connected by this vector are slightly more parallel than the corresponding regions for $r=1$, thus leading to better nesting and an increase in $\chi_s^{\rm RPA}(q,\omega=0)$ for $q=(\pi,\pi)$. The eigenfunction $g_d(k)$ (panel {\bf f}) changes slightly for $r=0.9$, but retains mostly its $d_{x^2-y^2}$ structure. Due to the breaking of C4 rotational symmetry, there is a small admixture of an $s$-wave component that manifests itself in a change in the location of the nodal points ($g_d(k)=0$) away from the diagonal line $k_x=k_y$ and also in a difference in the magnitude of $g_d(k)$ for $k$ near $(0,0)$ ($(\pi,0)$ in full Brillouin zone) and $k$ near $(0,\pi)$.

For the $r=0.25$ case of weakly coupled ladders (bottom panels {\bf g}-{\bf i}), the bandstructure has separated into bonding and anti-bonding bands, and the Fermi surface, as can be seen from panel {\bf i}, is becoming more one-dimensional. Nevertheless, the wave vector of peak intensity in $\chi^{\rm RPA}_s(q,0)$ remains near $q=(\pi,\pi)$, although it has become slightly incommensurate. The peak intensity is also reduced compared to $r=0.9$, but remains larger than for $r=1$. We believe that the reduced pairing strength $\lambda_d=0.16$ for this case is therefore due to the fact that the peak position in $\chi^{\rm RPA}_s(q,0)$ has moved away from $(\pi,\pi)$. Even in this limit, the leading eigenfunction $g_d(k)$ mostly retains its $d_{x^2-y^2}$ symmetry, but due to the structure of the Fermi surface becomes node-less. Moreover, it switches sign between the bonding (outer, purple) and anti-bonding (inner, brown) Fermi surface segments, and shows very little variation in magnitude along these segments, especially along the anti-bonding segments. 

We now use an RPA/BCS formalism \cite{Eremin2005,Norman2007,Maier2008,Maier2009} to study the magnetic spin excitation spectrum in the superconducting state as the hopping ratio $r$ is tuned away from the uniform limit. For the uniform $r=1$ case, this formalism is well known to give a resonance in $\chi^{\rm RPA''}_s(q,\omega)$ \cite{Eremin2005,Norman2007} and thus provides a framework in which one can understand the superconducting state neutron resonance that is found in the doped cuprates \cite{Scalapino2012}. This resonance in $\chi^{\rm RPA''}_s(q,\omega)$ is generally found for a wave-vector $q$ that connects regions on the Fermi surface between which the superconducting gap $\Delta(k)$ changes sign, i.e. $\Delta(k+q) = -\Delta(k)$, and therefore has been used as clear evidence for an unconventional superconducting state \cite{Scalapino2012}. Here, we are studying how this neutron resonance evolves as the hopping parameter $r$ is varied and the system changes from the uniform $r=1$ limit to an array of weakly coupled Hubbard ladders. Since we generally find eigenfunctions (gap structures) $g_d(k)$ that change sign on the Fermi surface for all values of $r$, we expect a resonance to occur for all $r$. However, the changes in the Fermi surface under variations of $r$ may significantly affect the wave-vector $q$ at which this resonance appears, as well its energy ($\omega$) dispersion as $q$ is varied.

In the superconducting state, the expression for $\chi^0$ in Eq.~\eqref{eq:chi0} changes and aquires an additional term from the anomalous component $F^0$ of the Green's function
\begin{eqnarray}\label{eq:chi0F}
  \chi^0_{\ell_1\ell_2\ell_3\ell_4}(q,i\omega_m)&&=-\frac{T}{N}\sum_{k,\omega_n}
  \big\{G^0_{\ell_3\ell_1}(k+q,i\omega_n+i\omega_m) \\\nonumber
  &&\hspace{2cm}\times \,G^0_{\ell_2\ell_4}(k,i\omega_n)\\ \nonumber
  &&\hspace{2cm} + F^0_{\ell_1\ell_3}(-k-q,-i\omega_n-i\omega_m)\\\nonumber
  &&\hspace{2cm} \times F^0_{\ell_2\ell_4}(k,i\omega_n)\big\}
\end{eqnarray}
with
\begin{eqnarray}
    G^0_{\ell\ell'}(k,i\omega_n) &=& \sum_\mu a^\ell_\mu(k)a^{\ell^\prime *}_\mu(k) \frac{i\omega_n+\xi_\mu(k)}{\omega_n^2+E^2_\mu(k)}\\
    F^0_{\ell\ell'}(k,i\omega_n) &=& \sum_\mu a^\ell_\mu(k)a^{\ell^\prime}_\mu(k) \frac{\Delta_\mu(k)}{\omega_n^2+E^2_\mu(k)}\,.
\end{eqnarray}
Here $E_\mu(k) = \sqrt{\xi_\mu^2(k)+\Delta_\mu^2(k)}$ and $\Delta_\mu(k)$ is the superconducting gap on band $\mu$. To determine $\Delta(k)$ for all $k$ in the Brillouin zone, one could take the RPA results for the leading gap eigenfunction $g_d(k)$ and parametrize it in terms of crystal harmonics \cite{Maier2009,Maier2012}. However, since $g_d(k)$ is only known for $k$ on the Fermi surface, this procedure can lead to non-unique parametrizations. In fact, for the system with $r<1$, we have indeed encountered difficulties in finding unique solutions. This is possibly due the fact that C4 symmetry is lost and other $s$-wave crystal harmonics contribute and the absence of orthogonality of different crystal harmonics for $k$ only on the Fermi surface. 

\begin{figure*}[t!]
    \centering
    \includegraphics[width=0.93\textwidth]{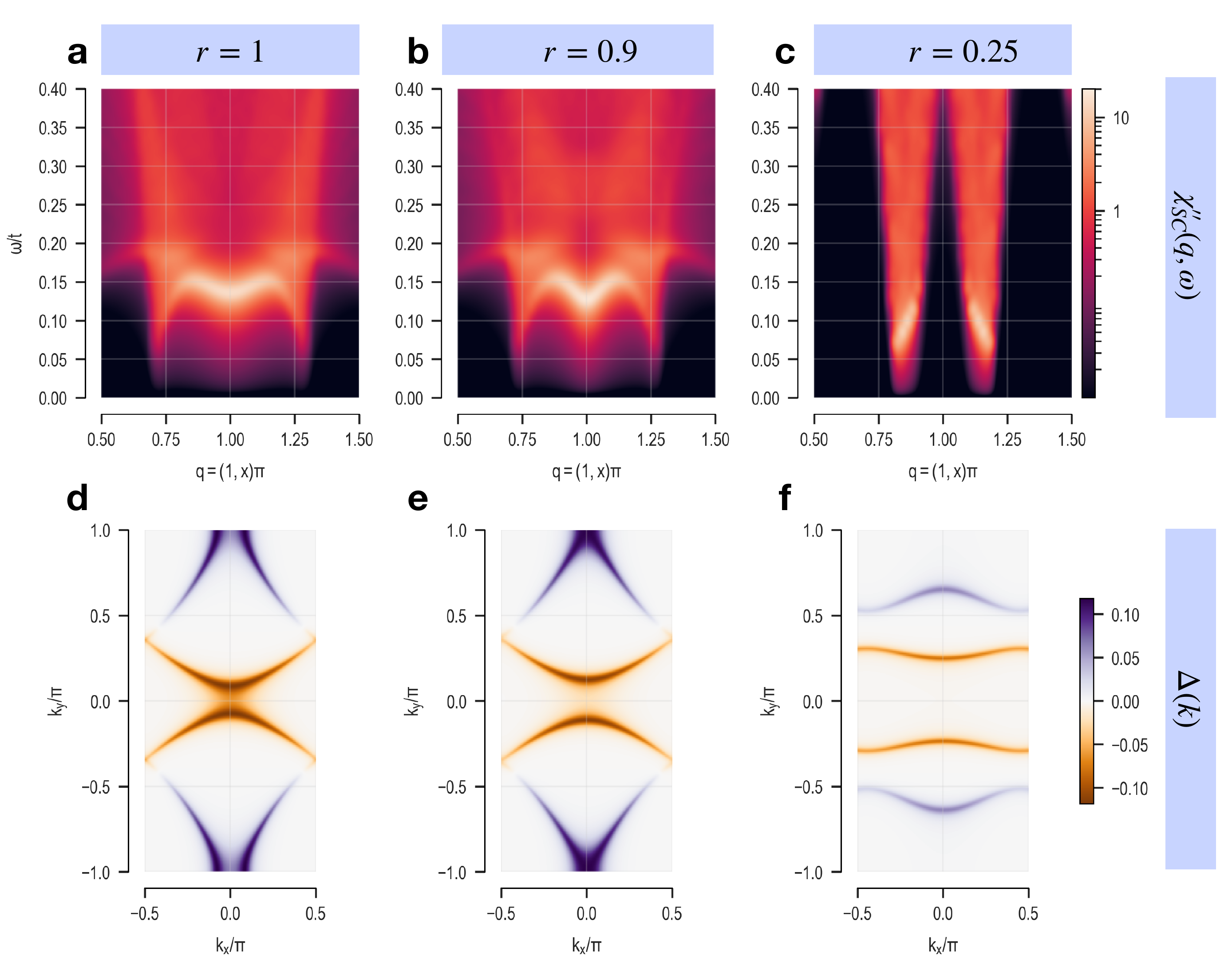}
    \caption{Imaginary part of the RPA/BCS dynamic spin susceptibility $\chi''_{SC}({\bf q}, \omega)$ in the superconducting state along ${\bf q}=(\pi,0.5\pi)$ – $(\pi,1.5\pi)$ for the Hubbard model with $\langle n\rangle=0.85$ and $U=1.8t$ for different values of the hopping ratio $r$ (panels {\bf a} - {\bf c}). The superconducting gap structures shown in panels {\bf d} - {\bf f} were used. For the uniform case ($r=1$), the spectrum displays the usual hour glass dispersion with a minimum in the resonance energy at ${\bf q}=(\pi,\pi)$. For the weakly coupled ladders ($r=0.25$), a resonance appears at incommensurate wave-vectors that disperses upward as ${\bf q}$ moves toward $(\pi,\pi)$.
    \label{fig:chiw}} 
\end{figure*}

We therefore take a different approach and fix $\Delta_\mu(k, r)$ using the following procedure, which defines the gap $\Delta_{\ell\ell'}(k)$ first in orbital space: Considering only nearest-neighbor pairing, we set $\Delta_{11/22}(k_y) = \Delta_0\cos k_y$ along the leg-direction in the ladder. Along the rung-direction, within the ladder, we set $\Delta_{12/21} = -\Delta_0$ and between adjacent ladders $\Delta_{12/21}=-r\Delta_0$, resulting in $\Delta_{12}(k_x) = \Delta^*_{21}(k_x)=0.5e^{ik_x}+0.5re^{-ik_x}$. This reduces to the usual $d_{x^2-y^2}$-wave $\Delta_0(\cos k_x-\cos k_y)$ gap for the uniform $r=1$ case. In the opposite $r=0$ limit, it results in a gap that is only finite within the ladders, $\Delta_0$ along the leg direction, $-\Delta_0$ along the rung direction. We then transform $\Delta_{\ell\ell'}(k)$ to the band representation to give 
\begin{equation}
    \Delta_\mu(k) = \sum_{\ell\ell'} a_\mu^\ell(k) \Delta_{\ell\ell'}(k)a_\mu^{\ell'}(-k)\,.
\end{equation}
In addition, in order to describe the behavior of the order parameter away from the Fermi surface, we have multiplied $\Delta_\mu(k)$ by a Gaussian cut-off $\exp\{-[\xi_\mu(k)/\Omega_0]^2\}$ with $\Omega_0=0.1t$.  The gap magnitude is set to $\Delta_0=0.06t$ for all calculations, resulting in an anti-nodal gap magnitude of $2\Delta_0=0.12t$ for the uniform case. The resulting gap $\Delta_\mu(k)$ in band representation is shown in Fig.~\ref{fig:chiw}, panels {\bf d}-{\bf f} for $r=1$, $0.9$ and $0.25$, respectively. Comparing with panels {\bf c}, {\bf f} and {\bf i} in Fig.~\ref{fig:Ek_chi_g}, one sees that this gap structure represents the leading RPA eigenfunction $g_d(k)$  well, including in particular the rather isotropic nature of the nodeless $r=0.25$ case.

Using this gap structure, we have calculated the dynamic spin susceptibility $\chi^{RPA}_s(q,\omega)$ (Eq.~\eqref{eq:dynamicSpinSus}) using Eqs.~\eqref{eq:chi} and \eqref{eq:chi0F} for different $r$. The top panels {\bf a}-{\bf c} in Fig.~\ref{fig:chiw} show the imaginary part of $\chi^{RPA}_{SC}(q,\omega)$ calculated in the superconducting state for $q$ along the line from $(\pi,0.5\pi)$ to $(\pi,1.5\pi)$. For the uniform $r=1$ case, we find the usual hour-glass like spectrum \cite{Eremin2005,Norman2007}. In the superconducting state, a spin gap opens and low energy normal state spectral weight is transferred to a resonance at $\omega\approx 0.15t$, i.e. inside the spin gap $2\Delta_0=0.24t$. The resonance energy has a minimum at $q=(\pi,\pi)$ and disperses upwards on either side. For the slightly modulated $r=0.9$ case, the spectrum is very similar with the resonance becoming slightly stronger and moving down in energy. This is due to the enhanced nesting for this case as discussed already in the context of Fig.~\ref{fig:Ek_chi_g}. 

In the weakly coupled ladder case, $r=0.25$, the spectrum changes drastically and displays two columns of scattering centered at incommensurate $q$-vectors. The resonance completely disappears at $q=(\pi,\pi)$. At smaller (and larger) $q_y\approx 0.85$ ($1.15$) a strong resonance is seen that displays a downward dispersion as $q$ moves away from $(\pi,\pi)$. This result is consistent with the static spin susceptibility shown for this case in Fig.~\ref{fig:Ek_chi_g} {\bf h}, where the peak of highest intensity has moved away from $q=(\pi,\pi)$ to smaller $q_y$. For this case, $q=(\pi,\pi)$ does not connect different pieces of the Fermi surface. Rather, they are connected by $q=(\pi,q_y)$ with $q_y$ reduced to $\sim 0.85\pi$. As $q_y$ decreases further, $(\pi,q_y)$ connects regions on the Fermi surface where the gap $\Delta_\mu(k)$ has decreased, explaining the downward dispersion of the resonance with decreasing $q_y$. Compared to the cases with $r\lesssim 1$, the resonance for $r=0.25$ appears at much smaller energies. This is due to the fact that the gap $\Delta_\mu(k)$ (panel {\bf f}) is smaller for this case. 

\section{Summary and Conclusions}
We have used a weak-coupling RPA spin-fluctuation formalism to analyze the superconducting pairing properties of a two-dimensional Hubbard model with modulated hopping amplitudes as it evolves from the uniform square lattice limit to an array of weakly coupled two-leg ladder. For the values of the inter-ladder to intra-ladder hopping ratio $r$ we have studied, we have found a leading pairing instability with dominant $d_{x^2-y^2}$ structure. The $d$-wave pairing strength $\lambda_d$ is found to have a maximum for $r=0.9$ representing a system with a slight unidirectional modulation in the hopping amplitudes. This enhancement in the pairing correlations over the uniform 2D limit arises from an increase in the spin-fluctuation pairing interaction due to increased nesting in the Fermi surface. 

We then calculated the dynamical magnetic susceptibility in the superconducting state using an RPA/BCS formalism. We have shown that the resonance that is found in the uniform model for commensurate $q=(\pi,\pi)$ becomes {\it stronger} for $r=0.9$, and then evolves into two distinct resonances at incommensurate wave-vectors for $r=0.25$ when the two-leg ladders are weakly coupled. These results provide new insight into the general question of how the electronic structure can be changed to optimize superconductivity, and how these changes in the electronic structure, and the resulting pairing properties, manifest themselves in the magnetic excitation spectrum that can be measured in inelastic neutron scattering experiments. 

We conclude reminding readers that the study of ladders has received renewed attention in the context of iron-based superconductors where it has been reported that the ladder materials BaFe$_2$X$_3$ (X = S, Se) become superconducting at high pressure~\cite{NatMatSC,PRLSC,lei,Dagotto:Rmp,patel,yangpressure1,yangpressure2,Zhang:prb20-2,nocera}, similarly as the Cu-ladders do. Moreover, exotic spin states involving antiferromagnetically coupled ferromagnetic spin ``blocks'' have been studied experimentally and theoretically~\cite{rincon,caron1,caron2,mourigal,wang2016,yang1,yang2,jacek}, and predictions for such unusual spins arrangement in diffraction neutron scattering were made and confirmed.
This is a fertile area where inelastic neutron scattering can make an impact similar as in cuprates when single crystals become available. Predicting their properties within RPA in the unusual $s\pm$ pairing state or in the doped block states for an array of weakly coupled iron ladders is a challenge to be addressed in the near future.

\begin{acknowledgements}
This work was supported by the U.S. Department of Energy, Office of Basic Energy Sciences, Materials Sciences and Engineering Division. This manuscript has been authored by UT-Battelle, LLC, under Contract No. DE-AC0500OR22725 with the U.S. Department of Energy. The United States Government retains and the publisher, by accepting the article for publication, acknowledges that the United States Government retains a nonexclusive, paid-up, irrevocable, world-wide license to publish or reproduce the published form of this manuscript, or allow others to do so, for the United States Government purposes. The Department of Energy will provide public access to these results of federally sponsored research in accordance with the DOE Public Access Plan (\url{http://energy.gov/downloads/doe-public-access-plan}).
\end{acknowledgements}

\bibliography{main}

\end{document}